\documentclass[10pt,twocolumn,letterpaper]{article}

\usepackage[pagenumbers]{iccv} 
\usepackage{multirow}

\definecolor{iccvblue}{rgb}{0.21,0.49,0.74}
\usepackage[pagebackref,breaklinks,colorlinks,allcolors=iccvblue]{hyperref}

\def\paperID{13240} 
\def\confName{ICCV}
\def\confYear{2025}

\title{AdaViT: Adaptive Vision Transformer for Flexible Pretrain and Finetune with Variable 3D Medical Image Modalities}

\author{
    Badhan Kumar Das$^{\ddagger \S}$\thanks{Equal Contribution} \qquad  
    Gengyan Zhao$^{\dagger}$\textsuperscript{*} \qquad  
    Han Liu$^{\dagger}$ \qquad  
    Thomas J. Re$^{\dagger}$ \\ 
    Dorin Comaniciu$^{\dagger}$ \qquad  
    Eli Gibson$^{\dagger}$ \qquad  
    Andreas Maier$^{\S}$
    \\
    $^{\ddagger}$ Siemens Healthineers AG \\
    $^{\dagger}$ Siemens Medical Solutions USA, Inc. \\
    $^{\S}$ FAU Erlangen-Nuremberg
}

\begin{document}
\maketitle
\begin{abstract}

Pretrain techniques, whether supervised or self-supervised, are widely used in deep learning to enhance model performance. In real-world clinical scenarios, different sets of magnetic resonance (MR) contrasts are often acquired for different subjects/cases, creating challenges for deep learning models assuming consistent input modalities among all the cases and between pretrain and finetune. Existing methods struggle to maintain performance when there is an input modality/contrast set mismatch with the pretrained model, often resulting in degraded accuracy. We propose an adaptive Vision Transformer (AdaViT) framework capable of handling variable set of input modalities for each case. We utilize a dynamic tokenizer to encode different input image modalities to tokens and take advantage of the characteristics of the transformer to build attention mechanism across variable length of tokens. Through extensive experiments, we demonstrate that this architecture effectively transfers supervised pretrained models to new datasets with different input modality/contrast sets, resulting in superior performance on zero-shot testing, few-shot finetuning, and backward transferring in brain infarct and brain tumor segmentation tasks. Additionally, for self-supervised pretrain, the proposed method is able to maximize the pretrain data and facilitate transferring to diverse downstream tasks with variable sets of input modalities.

\end{abstract}    
\section{Introduction}
\label{sec:intro}

In recent years, pretrain techniques have become essential in advancing deep learning. Both supervised and self-supervised pretrain methods are commonly used to initialize models with rich feature representations, thereby enhancing their performance on downstream tasks \cite{chen2020simple,yosinski2014transferable,he2020momentum}. By learning generalizable patterns in the pretrain stage, models can more easily adapt to specific applications when finetuned on task-specific datasets. These approaches have demonstrated remarkable success across various domains, including natural language processing \cite{kenton2019bert}, computer vision \cite{krizhevsky2012imagenet,chen2020simple}, and, increasingly, medical imaging \cite{irvin2019chexpert,zhou2019models,zhou2023self}.

\begin{figure}[htb!]
    

  \centering
  \centerline{\includegraphics[width=1\linewidth]{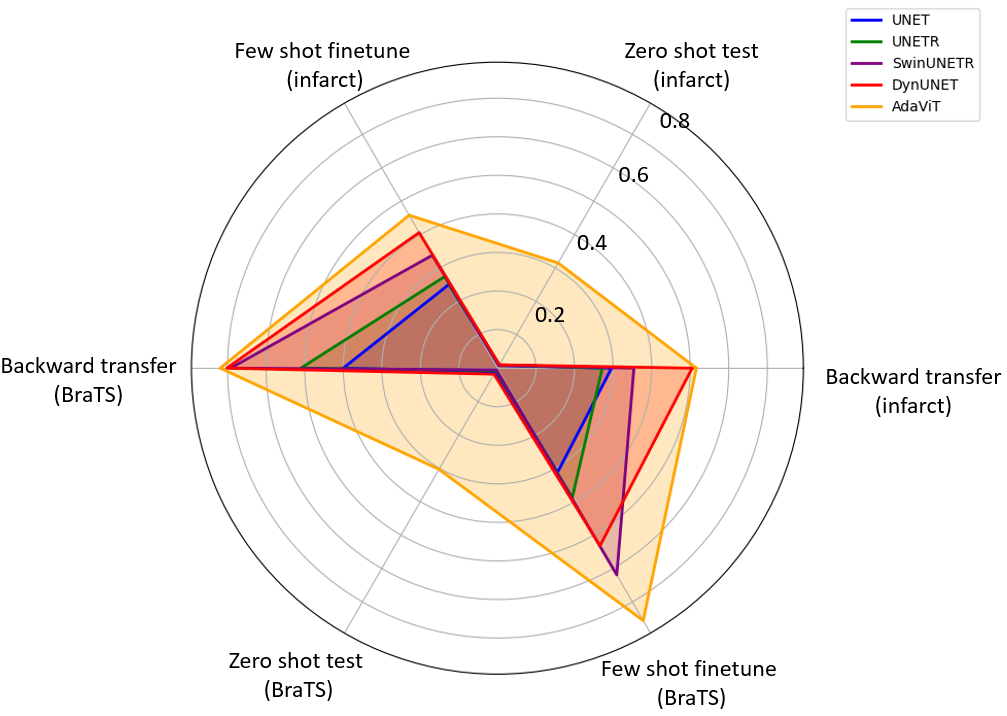}}
\caption{Performance comparison of different models on brain infarct and brain tumor segmentation in zero-shot testing, few-shot finetuning, and backward transferring.}
\label{compare_chart}
\end{figure}

In medical imaging, accurately diagnosing a disease or segmenting a lesion often requires multiple image modalities of the same subject/case. This poses challenges to the current pretrain-finetune framework, since each subject/case may have a different set of imaging modalities, such as various magnetic resonance (MR) contrasts. This is due to different acquisition protocols from different hospitals or clinical tasks, resulting in a wide range of input variations \cite{wen2021rethinking,yu2022transfer,kora2022transfer}. In supervised pretrain, when a model is trained on a specific set of modalities for a given task, it may struggle when finetuned on a different set of modalities for the same task \cite{guan2021domain,ben2006analysis}, highlighting the need for an adaptable model capable of handling variable input modalities across both stages. Such a model would maintain strong performance even with mismatched input modality sets between pretrain and finetune, thereby extending pretrain’s benefit to a wider range of downstream tasks without the constraint on input modality set consistency. Moreover, such an adaptable model can also maximize the data availability during both pretraining (supervised or self-supervised) and finetuning, since it can handle cases with variable sets of input modalities from iteration to iteration. Therefore, the cases with a different set of modalities from the majority don’t need to be removed. This need of adaptability is also prevalent in broader topics, including foundation models \cite{azad2023foundational}, federated learning \cite{li2021survey}, and continual learning \cite{wang2024comprehensive}, where diverse and evolving heterogeneous data inputs must be taken care of.


To address these challenges, we propose an adaptive Vision Transformer (AdaViT) framework capable of processing a variable set of input image modalities from each case. Our framework employs a dynamic tokenizer to generate tokens from each case’s each input image modality, and a transformer encoder to process the variable-length long sequence composed of all the tokens from all the modalities of each case. This architecture enables the model to transfer pretrained knowledge effectively to new datasets with different input modality types, achieving excellent performance on downstream tasks. Through extensive tests, we show that the model achieves superior performance on zero-shot testing, few-shot finetuning, and backward transferring in brain infarct and brain tumor segmentation tasks, as shown in Figure 1. Furthermore, by permitting different set of input modalities for each case, our approach optimizes data use in the context of supervised and self-supervised pretrain and facilitates adaptability to a variety of downstream tasks, enabling more flexible and heterogeneous clinical imaging applications.

We summarize our main contributions as follows:

\begin{itemize}
\item AdaViT Framework: We propose a novel framework that enables flexible pretrain and finetune for both supervised and self-supervised methods with a variable set of 3D medical image modalities for each case, addressing the challenges posed by data heterogeneity in medical imaging. 
\item Supervised Pretrain and Finetune: Our framework outperforms conventional methods in zero-shot test, fewshot learning and backward transferring scenarios, adapting from pretrain to finetune datasets with varying input modality sets. The key objective is to efficiently transfer knowledge learned from the pretrain dataset to the finetune dataset, improving the model’s flexibility and performance in handling diverse input conditions across cases and between pretrain and finetune.
\item Self-supervised Pretrain and Finetune: We extend our approach to self-supervised learning, where by accommodating a variable number of 3D input modalities our model architecture maximizes data availability and model generalization during pretraining and extends its utility across a broad spectrum of downstream tasks. Consequently, the performance of downstream tasks is improved while the associated effort is minimized. 

\end{itemize}

\section{Related Work}

\subsection{Multi-modality in Medical Imaging}
In medical imaging, accurate diagnosis or segmentation usually requires multiple image modalities. Especially for magnetic resonance imaging (MRI), usually multiple MR images with different contrasts are acquired for the same subject in a single acquisition session. These MR image modalities are considered as “paired” data and are typically used simultaneously, since they have the same aligned structure of the same subject but different contrasts to capture different characteristics of the same object. This is similar to the multi-spectral image \cite{zhang2021guided} or RGB-depth image \cite{giannone2019learning} in computer vision, but is different from the multi-modality data in other non-medical field, where the modalities (e.g., images, audios) often have different objects/contents (“unpaired”) and are typically processed by the model separately \cite{zhang2023meta,xie2022unimiss,girdhar2022omnivore}.

For real-world clinical MR data, different subjects/cases/studies usually consist of different sets of MR contrasts due to different acquisition protocols or tasks at different clinical cites. This poses challenges for the current deep learning methods to handle large scale heterogeneous datasets, and to transfer the pretrained model to different downstream tasks with different set of input modalities, since these methods typically require all the input subjects/cases have a fixed set of image modalities/contrasts. The novelty of this work is to handle different combinations of “paired” modalities with a flexible framework.

\subsection{Supervised Pretrain and Finetune}
Supervised learning (SL) aims to reduce the discrepancy between AI predictions and human-annotated semantic labels, typically by learning robust feature representations from the same or similar tasks. Historically, supervised pretrain on large-scale datasets like ImageNet \cite{deng2009imagenet} has demonstrated substantial success in transfer learning \cite{yosinski2014transferable}. The effectiveness of transfer learning can further increase when models are supervised pretrained on even larger natural image datasets, including ImageNet-21K \cite{kolesnikov2020big}, Instagram \cite{mahajan2018exploring} and JFT-3B \cite{zhai2022scaling}. 

When sufficient annotated data is available, supervised pretrain generally offers performance advantages over self-supervised methods \cite{steiner2021train, ridnik2021imagenet}, as it provides more direct learning from human-labeled semantics. 
Supervised pretrain on natural images has proven advantages for 2D medical imaging tasks following transfer learning \cite{shin2016deep,zhou2017fine}. 
Tajbakhsh et al. \cite{tajbakhsh2016convolutional} demonstrated that finetuning natural-image pretrained CNNs on medical imaging tasks can either outperform or match models trained from scratch and exhibit enhanced robustness to variations in dataset size. 
Supervised pretrain on 3D medical images has also been proven effective on 3D downstream medical imaging tasks. Zhou et al. \cite{liwell} applied it to both CNN and transformer based 3D models using the AbdomenAtlas 1.1 dataset \cite{li2024abdomenatlas} (CT images) for supervised pretrain and achieved excellent finetune performance on the TotalSegmentator v1 dataset \cite{wasserthal2023totalsegmentator} as well as a proprietary dataset.

However, Moatti et al. \cite{moatti2022domain} observed that conventional methods, including UNETR \cite{hatamizadeh2022unetr} and SwinUNETR \cite{hatamizadeh2021swin}, struggle when the input modality set varies. Existing deep learning models typically struggle when used for supervised pretrain on a specific task with a set of modalities and then later finetuned with a different set of modalities. This limitation arises because these models are often designed with the assumption that the input modalities remain consistent between pretrain and finetune for that task, leading to decreased performance when the input modality configuration changes.

We propose the AdaViT framework capable of handling a variable set of input modalities from each case, effectively addressing modality set mismatch between supervised pretrain and finetune. By flexibly accommodating available modalities for each case, it both maximizes data utilization during pretraining and enables successful weight transfer for finetuning on new datasets with different modality sets. This framework demonstrates robust performance in zero-shot testing and few-shot finetuning, even when the input modality set varies. Furthermore, it excels in backward transfer, ensuring consistent performance across segmentation tasks in different training stages on different datasets with different input modalities.

\subsection{Self-supervised Pretrain and Finetune}
Self-supervised learning (SSL) shows success in building large-scale deep learning-based applications by leveraging a large amount of unlabeled data. Early works in this field developed tailored pretext tasks that mainly aim to reconstruct images from the original or distorted images and in the process learn meaningful features. 
For example, Zhou et al. \cite{zhou2019models} used an autoencoder to reconstruct images with non-linear distortions, inpainting, and local shuffling, and transfer the learned knowledge to classification and segmentation tasks on CT and X-ray images.
Chen et al. \cite{chen2020simple} proposed a context restoration task to enhance model robustness in various medical applications by generating distorted images with preserved intensity through random patch swapping within the input.
Taleb et al. \cite{taleb20203d} extended this line of work by introducing the multimodal puzzle task, inspired by Jigsaw puzzles, to leverage representation learning across multiple medical image modalities.  

In recent years, contrastive learning has provided a more flexible approach for feature learning in medical imaging \cite{tang2022self,dippel2021towards}. For example, Sowrirajan et al.\cite{sowrirajan2021moco} leveraged MoCo \cite{he2020momentum} to pretrain on the large-scale unlabeled CheXpert dataset \cite{irvin2019chexpert} and then finetuned on the Shenzhen Hospital X-ray dataset \cite{jaeger2014two} to detect pleural effusion, demonstrating the utility of contrastive learning for transferring learned representations. 

Typically, self-supervised methods assume that each case has the same set of input modalities and the same set of input modalities is used in both the pretrain and finetune. However, in real-world clinical scenarios, each case can have a different number of MR modalities based on the clinical setup and protocols. Our proposed architecture is able to use all the available modalities from all the available cases. This maximizes the data availability for training and can fulfill the large-scale of data needed for self-supervised pretrain. Additionally, our pretrained model can be adapted to a range of different downstream tasks with varying input modality requirements, enabling us to ignore any specific modality constraints during pretrain.

\section{Method}

\begin{figure*}[htb!]
    

  \centering
  \centerline{\includegraphics[width=1.\linewidth]{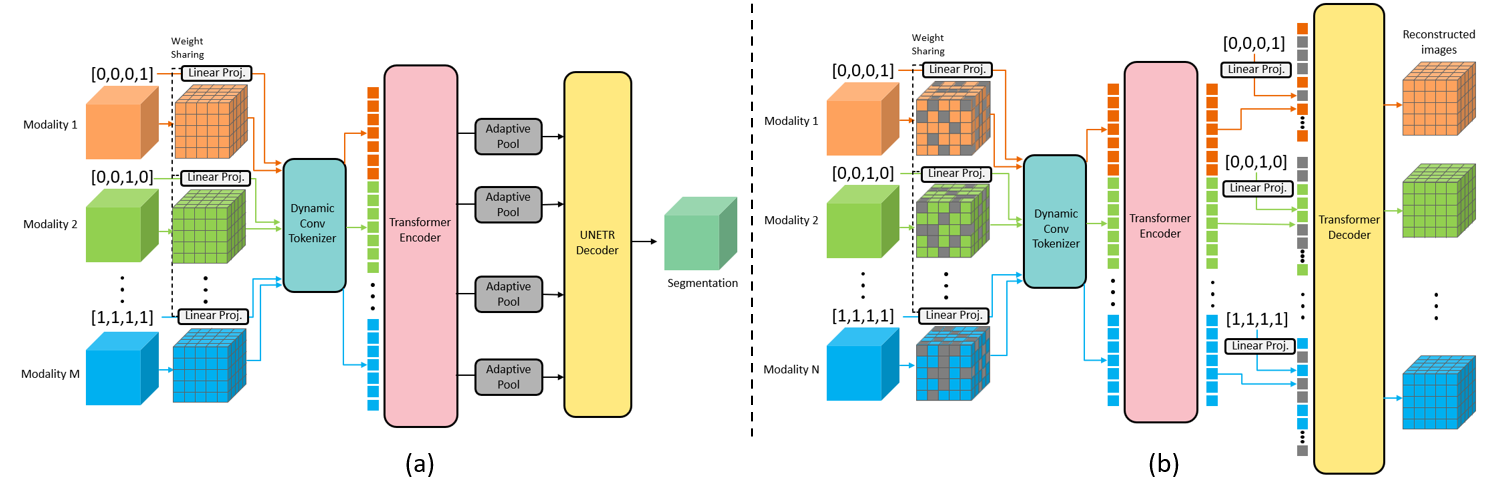}}
\caption{(a): Overview of AdaViT for supervised pretrain and finetune for segmentation task. (b): AdaViT in a masked autoencoder setting for self-supervised pretraining. Both architectures can handle variable set of input modalities from each case.}
\label{architrecture}
\end{figure*}

    


\subsection{Adaptive Vision Transformer}

 Similar to the original Vision Transformer (ViT) \cite{dosovitskiy2020image}, 
 3D non-overlapping patches are tokenized from 3D input images. However, in our framework, the different image modality/contrast volumes from the same case are not concatenated along the channel dimension. The patches of each input volume of the same case are tokenized separately. Then all the tokens of all the image modalities/contrasts of the same case are concatenated to form a long sequence to be sent to the transformer for encoding by a global self-attention.
 
First, inspired by the Dynamic Filter Network used in ModDrop++ \cite{liu2022moddrop++}, we propose a 3D Dynamic Convolution Tokenizer (DCT). The DCT is designed to dynamically adapt to different input modalities. Each modality is assigned a unique modality vector  \( \mathbf{m}_i \) of length \( l \), where \( i \) represents the $i$-th modality:


\[
\mathbf{m}_i \in \mathbb{R}^l
\]

The modality vector \( \mathbf{m}_i \) is passed through a learnable linear projector and then split to generate a weight vector \( \mathbf{w_i}_{mod} \) and a bias vector \( \mathbf{b_i}_{mod} \). The weight vector \( \mathbf{w_i}_{mod} \) and bias vector \( \mathbf{b_i}_{mod} \) are used to update the weights and biases of the dynamic convolutional layer. In this way, the convolutional kernels can learn to adjust their behavior when handling different input modalities to extract the most effective features from each modality \cite{yang2021unified,zhang2021dodnet,hu2022domain}. Let \( \mathbf{W} \) and \( \mathbf{B} \) denote the original weights and biases of the convolutional layer. The updated weights \( \mathbf{W_i}_{updated} \), biases \( \mathbf{B_i}_{updated} \) and the output tokens \( \mathbf{Y}_{i} \) from the $i$-th input modality/contrast \( \mathbf{X}_{i} \) are:

\[
\mathbf{W_i}_{updated} = \mathbf{W} \cdot \mathbf{w_i}_{mod}
\]

\[
\mathbf{B_i}_{updated} = \mathbf{B} \cdot \mathbf{b_i}_{mod}
\]


\[
\mathbf{{Y}_i} = \text{Conv}(\mathbf{X_i}, \mathbf{W_i}_{updated}, \mathbf{B_i}_{updated})
\]



Second, we take advantage of Transformer’s ability of processing variable length of input sequence of tokens to use it as an encoder to 1) handle variable sets of input modalities/contrasts from each case; 2) to learn the relationship of the modalities/contrasts from the same case from a global view. The tokens from all the modalities of the same case are concatenated into a long sequence and then fed into the transformer encoder. Each token receives a positional embedding (either sinusoidal or learnable) based on its 3D patch’s location in the 3D volume. Two tokens from different modalities but the same patch location are encoded by DCT’s different kernels controlled by the modality vector, but will have the same positional embedding, which helps preserving the spatial relationship of the tokens. The transformer block of the encoder applies a global self-attention across all the tokens from all the available modalities of each case (the number of available modalities and the modality types can be different from case to case). In this way, the inter-modality relationship and the global features from all the modalities can be learned, while the input modality set mismatch among cases or between pretrain and finetune can be solved as well.

\subsection{Supervised Pretrain and Finetune}

AdaViT for segmentation architecture is shown in Figure \ref{architrecture}(a). In supervised pretrain the proposed architecture is first trained with the pretrain dataset,  and later the pretrained weights are finetuned with the finetune dataset.

For supervised pretrain and finetune in segmentation tasks, we integrate the UNETR decoder head into the proposed AdaViT, where the DCT and the transformer encoder are used with a UNETR-based segmentation decoder head.

As we may have a variable number of inputs for each case our number of feature tokens also varies. We use an adaptive max pooling layer to fuse the extracted features from all the modalities at each level in the UNETR decoder head. In this way, the variable number of feature tokens can be transferred to tensors with fixed sizes at different decoding levels to generate segmentation masks.

Let \( \mathbf{X} \) denote the output of the transformer encoder:

\[
\mathbf{X} \in \mathbb{R}^{B \times  \left(N \times \frac{H \times W \times D}{\text{Patchsize}^3}\right) \times \text{EmbeddingDim}}
\]

where \( B \) is the batch size, \( N \) is the number of 3D input images, \(\left(\frac{H \times W \times D}{\text{Patchsize}^3}\right)\) is the number of the patches/tokens, \(\text{EmbeddingDim}\) is the transformer embedding dimension.

We use an adaptive max pooling layer to fuse extracted features from all tokens, resulting in:

\[
\mathbf{X}_{new} \in \mathbb{R}^{B \times \left(\frac{H \times W \times D}{\text{Patchsize}^3}\right) \times \text{EmbeddingDim}}
\]

\subsection{Self-supervised Pretrain}

The proposed adaptive framework employs a masked autoencoder (MAE) approach \cite{he2022masked} for self-supervised learning. The process begins with random masking off a percentage of 3D patches in each modality’s original 3D volume as shown in Figure \ref{architrecture}(b). The left unmasked patches are encoded to tokens using DCT as mentioned above. Then all the unmasked tokens from different modalities are concatenated to a long sequence and fed into the transformer encoder. As this encoder works only on unmasked tokens, it is possible to train with large embedding dimensions, large number of heads, or a large number of input image modalities from each case.

\begin{table*}[ht!]
\centering
\begin{tabular}{cccccc}
\hline
\textbf{Category} & \textbf{Dataset} & \textbf{Contrasts} & \textbf{Training} & \textbf{Validation} & \textbf{Test} \\ \hline
\multirow{4}{*}{\textbf{SL}} & SL-pretrain-Infarct & ADC, TraceW & 1556 & 184 & 158 \\ 
 & SL-finetune-Infarct & ADC, TraceW, T2 (Opt) & 92 & 9 & 48 \\ 
 & SL-pretrain-BraTS & FLAIR, T1CE & 700 & 100 & 125 \\ 
 & SL-finetune-BraTS & FLAIR, T1CE, T1 & 20 & 100 & 125 \\ \hline
\multirow{3}{*}{\textbf{SSL}} & SSL-pretrain & FLAIR, T1CE, T1, ADC, TraceW, T2, GRE, SWI & 45,374 & - & - \\ 
& SSL-finetune-Infarct & ADC, TraceW, T2 (Opt) & 1648 & 193 & 215 \\  
 & SSL-finetune-BraTS & FLAIR, T1CE, T1, T2 & 800 & 200 & 250 \\ \hline
\end{tabular}
\caption{Datasets for supervised pretrain and finetune (SL) and self-supervised pretrain and finetune (SSL). TraceW: Trace-weighted, ADC: Apparent Diffusion Coefficient, T2: T2-weighted, FLAIR: Fluid Attenuated Inversion Recovery, T1: T1-weighted, T1CE: post-contrast T1-weighted, GRE: Gradient Echo, SWI: Susceptibility-Weighted. Opt: optional, T2 may or may not be available for a case.}
\label{table:ssl_supervised_data}
\end{table*}

After getting the encoded output from the transformer encoder, a placeholder token is inserted at each masked token’s position. Both masked and unmasked tokens are then added with the positional and modality embedding again and fed into a transformer decoder with global self-attenstion across all the tokens from all the modalities of each case. A linear projector then maps the decoded tokens back to the size of the 3D patches to reconstruct the original 3D volumes of all the given input image modalities/contrasts.

\section{Experiments}

\subsection{Dataset Configuration}

We extensively evaluated our proposed AdaViT with two mainstream pretraining techniques: supervised pretrain and self-supervised pretrain, on 2 tasks: brain infarct and brain tumor segmentation. For brain infarct segmentation, we used an internal dataset with acute/subacute brain infarct segmentation masks. For brain tumor segmentation we used the BraTS 2021 dataset \cite{baid2021rsna} with segmentation masks of 3 tumor regions: tumor core (TC), whole tumor (WT), and enhancing tumor (ET). \textbf{For supervised pretrain and finetune}, we focus on the challenging yet practical real-world scenario where the input modalities between pretrain and finetune are mismatched. Thus, we split the cases in each dataset further into 2 subsets, and give the cases in each subset a unique set of “paired” input modalities, which is shown in Table \ref{table:ssl_supervised_data}. \textbf{For self-supervised pretrain and finetune:} For self-supervised pretrain, we utilize another extra-large unlabeled MR dataset of 45,374 cases. For each case, a variable number of MRI contrasts may be available. When any of the 8 contrasts shown in Table \ref{table:ssl_supervised_data} are available, we include all the available ones for each case. The self-supervised pretrained model is then finetuned on the brain infarct and tumor segmentation tasks, each requiring a different set of input MR contrasts. The detailed dataset configurations are also shown in Table \ref{table:ssl_supervised_data}. 

\subsection{Implementation}


For supervised pretrain and finetune,  we trained the models for 200 epochs and saved the weights with the best validation result for later use. The models were trained with a Dice loss, a weighted Adam optimizer, and a cosine annealing scheduler with an initial learning rate of 1e-4. For self-supervised learning pretrain, a weighted Adam optimizer, an L2 loss function, and a cosine annealing scheduler with an initial learning rate of 1e-5 were used. The SSL pretraining was done for 500 epochs with a 70\% masking ratio and a batch size of 4. The experiments were implemented with PyTorch (v1.12.1) and MONAI (v1.1.0) \cite{cardoso2022monai} frameworks. For segmentation tasks, we utilized the HGX-A100 system equipped with 4 A100 GPUs, 512 GB system memory, and 40 GB GPU memory per card. For self-supervised pretraining, we used HGX-A100 systems configured with 8 A100 GPUs, 1 TB system memory, and 40 GB GPU memory per card.

\subsection{Supervised Pretrain and Finetune}

For supervised pretrain and finetune, we first performed supervised pretrain and then performed 3 experiments on the pretrained model (zero-shot test, few-shot finetune, and backward transfer) on 2 tasks (brain infarct and tumor segmentation) to study how well the supervised pretrained models can handle input modality set mismatch in finetune.

\begin{table*}[h!]
\centering
\begin{tabular}{cccccc}
\hline
\multirow{2}{*}{\textbf{Model}} & \textbf{SL-finetune-Infarct} & \multicolumn{4}{c}{\textbf{SL-finetune-BraTS}} \\ \cline{3-6} 
 & \textbf{Mean Dice Score} & \textbf{TC} & \textbf{WT} & \textbf{ET} & \textbf{Average} \\ \hline
UNET & 0.009 & 0.004 & 0.021 & 0.008 & 0.011 \\ 
UNETR & 0.004 & 0.001 & 0.018 & 0.001 & 0.007 \\ 
SwinUNETR & 0.006 & 0.003 & 0.009 & 0.002 & 0.005 \\ 
DynUNET & 0.010 & 0.025 & 0.024 & 0.004 & 0.018 \\ 
AdaViT (ours) & \textbf{0.315} & 0.343 & 0.221 & 0.344 & \textbf{0.303} \\ \hline
\end{tabular}
\caption{Performance comparison of different models on zero-shot testing on SL-finetune-Infarct (mean Dice score) and SL-finetune-BraTS (Dice score on TC, WT, ET and the average Dice score) datasets.}
\label{table:inf_brats_finetune}
\end{table*}

First, we supervised pretrained the models using the training data in the SL-pretrain-Infarct and SL-pretrain-BraTS subsets respectively, and then performed zero-shot testing on the test data of the SL-finetune-Infarct and the SL-finetune-BraTS subsets where each case has a different set of contrasts comparing with the pretrain data sets. Second, we performed few-shot finetune on the supervised pretrained models with the training data in the SL-finetune-Infarct and SL-finetune-BraTS subsets to observe the models’ adaptability when input modalities set varied. Third, after finetuning, we also checked the finetuned models’ performance on the test data of the original pretrain subsets, SL-pretrain-Infarct and SL-pretrain-BraTS, to investigate if the models could retain the backward transfer performance on the original pretrain datasets after being finetuned with data having a different input modality set.

\textbf{Zero-shot test:} The zero-shot performance on the SL-finetune-Infarct and SL-finetune-BraTS test dataset is shown in Table \ref{table:inf_brats_finetune}. For the competing methods, the additional input modality is concatenated with the other inputs along the channel dimension, and the first convolutional layer’s input channel number is adjusted. The pretrained weights other than the new first convolutional layer are loaded. In the evaluations, the presence of additional input modalities presented challenges for these models, which struggled to handle a variable set of input modalities and produced notably low Dice scores. The proposed model achieved an average Dice score of 0.315 for infarct segmentation and 0.303 for brain tumor segmentation, outperforming all other compared segmentation methods. Its ability to maintain some segmentation performance given a different input modality set demonstrates its adaptability, which underscores AdaViT’s potential in clinical applications where imaging modalities/contrasts may vary widely from clinical site to clinical site, highlighting its advantage over conventional segmentation models.

\begin{table*}[htb!]
\centering
\begin{tabular}{cccccc}
\hline
\multirow{2}{*}{\textbf{Model}} & \textbf{SL-finetune-Infarct} & \multicolumn{4}{c}{\textbf{SL-finetune-BraTS}} \\ \cline{3-6} 
 & \textbf{Mean Dice Score} & \textbf{TC} & \textbf{WT} & \textbf{ET} & \textbf{Average} \\ \hline
UNET         & 0.296 & 0.452 & 0.663 & 0.086 & 0.400 \\ 
UNETR        & 0.272 & 0.498 & 0.607 & 0.424 & 0.510 \\ 
SwinUNETR    & 0.354 & 0.663 & 0.816 & 0.598 & 0.692 \\ 
DynUNET      & 0.505 & 0.594 & 0.807 & 0.705 & 0.702 \\ 
AdaViT (ours) & \textbf{0.516} & 0.673 & 0.764 & 0.722 & \textbf{0.720} \\ \hline
\end{tabular}
\caption{Performance comparison of different models in few-shot finetune on SL-finetune-Infarct (mean Dice score) and SL-finetune-BraTS (Dice score on TC, WT, ET and the average Dice score) datasets.}
\label{table:finetune}
\end{table*}

\textbf{Few-shot finetune:} After the few-shot finetune on the finetune datasets where each case has a different input modality set from the pretrain datasets, the model’s performance on the Infarct and BraTS finetune test datasets is presented in Table \ref{table:finetune}. In this scenario, AdaViT outperformed all other models, demonstrating its strong capacity for effective knowledge transfer to downstream tasks with limited training data and input modality set mismatch. For infarct segmentation, AdaViT achieved a mean Dice score of 0.516, surpassing all the other models. In brain tumor segmentation, AdaViT again achieved the highest average Dice score of 0.720, with high accuracy across the tumor core, whole tumor, and enhancing tumor regions. The few-shot learning performance suggests that AdaViT’s architecture is well-suited for rapid knowledge transfer in few-shot settings with variable multi-modal input configuration due to the flexibility of the model’s architecture design. This meets well the need in real-world clinical settings where 1) clinical data are expensive to acquire and annotate 2) data sharing is often limited by privacy policies 3) each case having a variable set of input modalities is common because of different acquisition protocols across different clinics. AdaViT’s effectiveness in few-shot scenarios highlights its significant potential for improving segmentation accuracy in these clinical situations, making it a valuable alternative to conventional segmentation models.

\begin{table*}[h!]
\centering
\begin{tabular}{cccccc}
\hline
\multirow{2}{*}{\textbf{Model}} & \textbf{SL-pretrain-Infarct} & \multicolumn{4}{c|}{\textbf{SL-pretrain-BraTS}} \\ \cline{3-6} 
 & \textbf{Mean Dice Score} & \textbf{TC} & \textbf{WT} & \textbf{ET} & \textbf{Average} \\ \hline
UNET         & 0.251 & 0.319 & 0.534 & 0.084 & 0.312 \\ 
UNETR        & 0.274 & 0.379 & 0.489 & 0.297 & 0.388 \\ 
SwinUNETR    & 0.338 & 0.577 & 0.743 & 0.536 & 0.619 \\ 
DynUNET      & 0.406 & 0.429 & 0.668 & 0.499 & 0.532 \\ 
AdaViT (ours) & \textbf{0.458} & 0.707 & 0.814 & 0.748 & \textbf{0.756} \\ \hline
\end{tabular}
\caption{Performance comparison of different models in backward transfer on SL-pretrain-Infarct (mean Dice score) and SL-pretrain-BraTS (Dice score on TC, WT, ET and the average Dice score) datasets.}
\label{table:pretrain}
\end{table*}

\textbf{Backward transfer:} 
After few-shot finetune, the models’ performances are tested on the test data of the original supervised pretrain subsets (SL-pretrain-Infarct and SL-pretrain-BraTS) to study the backward transfer performance. The results are shown in Table \ref{table:pretrain}. For the competing methods, the modality present in the finetune dataset but not in the pretrained dataset is set as an all-zero 3D volume. For infarct segmentation, AdaViT achieved the highest mean Dice score of 0.458, outperforming all other models, including SwinUNETR (0.338) and DynUNET (0.406). For brain tumor segmentation, AdaViT further excelled, achieving an average Dice score of 0.756 across tumor regions, with Dice scores of 0.707, 0.814, and 0.748 for the tumor core, whole tumor, and enhancing tumor, respectively. These results significantly outperform UNET and UNETR, which exhibited notable drops in backward transfer. AdaViT shows minimal performance degradation in both the infarct and brain tumor backward transfer tasks, highlighting its robustness with input modality set mismatch. This stability positions AdaViT as an excellent candidate for continual learning applications in multi-center or evolving clinical environments, where models need to adapt to variable input configurations among different clinics and maintain performance in backward transfer.

    



\begin{figure*}[ht!]
    

  \centering
  \centerline{\includegraphics[width=1.0\linewidth]{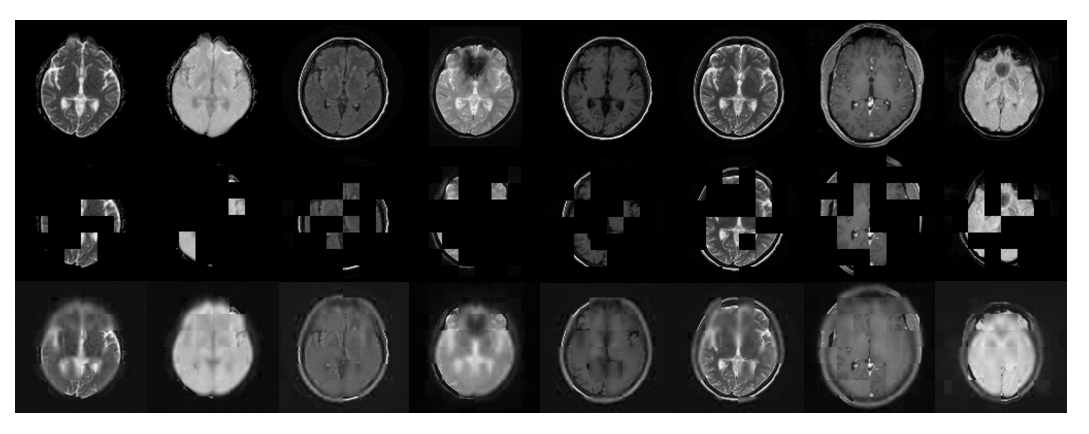}}
\caption{Reconstruction of self-supervised pretraining. First row: original image. Second row: masked image where masked patches are colored as black. Third row: reconstructed images. (Axial slices of ADC, TraceW, FLAIR, GRE, T1, T2, T1CE, and SWI are shown from left to right)}
\label{pretrain_rec}
\end{figure*}

From the results, it can be observed that AdaViT maintained high segmentation accuracy for both brain infarct segmentation and brain tumor segmentation in zero-shot testing, few-shot finetuning, and backward transfer. It significantly outperformed models like UNET, UNETR, and SwinUNETR when input modality set mismatch happens. The capacity to adjust to input modality set variation without drastic performance loss is valuable for real-world clinical scenarios, where data heterogeneity and inconsistency often arise due to different acquisition protocols at different clinical sites.

\subsection{Self-supervised Pretrain and Finetune}

\begin{table*}[ht!]
\centering
\begin{tabular}{cccccc}
\hline
\multirow{2}{*}{\textbf{Model}} & \multicolumn{1}{c}{\textbf{SSL-finetune-Infarct}} & \multicolumn{4}{c}{\textbf{SSL-finetune-BraTS}} \\ \cline{3-6}
 & \textbf{Mean Dice Score} & \textbf{TC} & \textbf{WT} & \textbf{ET} & \textbf{Average} \\ \hline

AdaViT & 0.561 & 0.758 & 0.852 & 0.782 & 0.797  \\ 
AdaViT with SSL & \textbf{0.598} & 0.778 & 0.853 & 0.811 & \textbf{0.814}  \\ \hline
\end{tabular}
\caption{Performance comparison of the AdaViT model without and with self-supervised pretrain being finetuned on the SSL-finetune-Infarct and SSL-finetune-BraTS datasets.}
\label{table:merged_results_infseg_brats}
\end{table*}

\begin{table*}[ht]
\centering
\begin{tabular}{ccccc}
\hline
\textbf{AdaViT Variants}  &\textbf{Embedding Dim} &\textbf{Num of Heads}&\textbf{Depth} & \textbf{Mean Dice Score} \\ \hline

AdaViT-base  & 768 & 12 & 12 & 0.598 \\ 
AdaViT-large & 1024 & 16 & 24 & 0.516 \\ 
AdaViT-huge & 1280 & 16 & 32 & 0.550 \\ \hline

\end{tabular}
\caption{Performance of AdaViT with self-supervised pretrain on brain infarct segmentation using different ViT sizes.}
\label{table:size}
\end{table*}

We further extended the AdaViT architecture for self-supervised pretrain using masked autoencoder. Our proposed AdaViT can accommodate variable set of input modalities for each case, enabling us to leverage all available cases with any sets of modalities for self-supervised pretrain. To investigate this, we first conducted self-supervised pretrain with our AdaViT on the SSL-pretrain dataset containing 45,374 cases, with up to 8 different modalities/contrasts for each case. We then finetuned the model for brain infarct and brain tumor segmentation tasks. The MAE style self-supervised pretrain reconstructions of ADC, TraceW, FLAIR, GRE, T1, T2, T1CE, and SWI with AdaViT are shown in Figure \ref{pretrain_rec}. The three rows consist of a single slice of the original input image of the above contrasts, their corresponding masked images, and the reconstructed images, respectively. The outcome shows that AdaViT masked autoencoders can recover randomly masked patches from the context. The recovered visible patches appear blurry due to L2 loss, which is a known phenomenon of MAE pretrain on 3D medical images \cite{zhou2023self}. In spite of the blurriness, the MAE reconstruction, as a self-supervised pretrain method, can learn to extract useful features benefiting the downstream tasks.

The effectiveness of the proposed AdaVIT with self-supervised pretrain is shown in Table \ref{table:merged_results_infseg_brats}. The self-supervised pretrain further enhanced model performance, where AdaViT with SSL outperformed the AdaViT without SSL by approximately 3.7\% for the infarct segmentation and by 1.7\% for brain tumor segmentation. This improvement highlights AdaViT’s ability to process variable set of input modalities for each case, which enables the model to maximize the data availability in self-supervised pretrain and to effectively transfer the learned knowledge from pretrain to downstream tasks despite variations in input modality requirements.

\subsection{Ablation Studies}

In our ablation studies, we examined the influence of varying ViT sizes with different embedding dimensions and number of attention heads in self-supervised pretrain and finetune on the brain infarct dataset. As shown in Table \ref{table:size}, AdaViT-base reaches th best performance which may be related to the size of the dataset used.

\begin{table}[ht]
\centering
\begin{tabular}{cc}
\hline
\textbf{Masking Ratio}  & \textbf{Mean Dice Score} \\ \hline

50\% & 0.516 \\ 
70\% & 0.598 \\ 

90\%  & 0.546\\ \hline


\end{tabular}
\caption{Performance of AdaViT with self-supervised pretrain using different masking ratios.}
\label{table:mask}
\end{table}

Additionally, we investigated the impact of different mask ratios on self-supervised pretrain for infarct segmentation, as summarized in Table \ref{table:mask}. 
A mask ratio of 70\% yielded the best performance in infarct segmentation, suggesting that a higher level of masking is beneficial for enhancing model robustness.

\begin{table}[h]
\centering
\begin{tabular}{cc}
\hline
\textbf{Method}  & \textbf{Mean Dice Score} \\ \hline

AdaViT with adaptive pooling & 0.598 \\ 
AdaViT with mean & 0.591 \\ 
\hline

\end{tabular}
\caption{Performance of AdaViT with adaptive pooling vs with mean on brain infarct segmentation}
\label{table:pool}
\end{table}

To integrate the UNETR decoder into our adaptive ViT framework we investigated two feature token fusion approaches to transfer the variable-length feature token sequence to a fixed-length sequence. One is using an adaptive pooling layer and the other is calculating the mean along the channel/modality dimension. The performance of these two approaches on brain infarct segmentation is shown in Table \ref{table:pool}.

\section{Conclusion}


In conclusion, we propose the AdaViT framework which can effectively accommodate variable sets of input modalities across different cases and between pretrain and finetune. It consistently outperforms other methods in zero-shot test, fewshot finetune and backward transfer when there is input modality set mismatch between supervised pretrain and finetune. Its strong flexibility enables it to maximize the data availability in self-supervised pretrain, and to efficiently transfer the learned knowledge to downstream tasks despite variations in input requirements, which further improves its performance. These characteristics make it well-suited for weight sharing in real-would clinical settings where highly diverse input modality availability across clinical sites or tasks may be an issue for conventional methods. AdaViT’s flexibility and robustness also show its promising potential for continual learning, federated learning and as a foundation model in clinical applications with heterogeneous data settings.
{
    \small
    \bibliographystyle{ieeenat_fullname}
    \bibliography{refer}
}

\end{document}